\documentclass[epj]{webofc}
\usepackage[utf8]{inputenc}
\usepackage[varg]{txfonts}   
\usepackage{booktabs}
\usepackage{xcolor}
\definecolor{darkred}{rgb}{0.4,0.0,0.0}
\definecolor{darkgreen}{rgb}{0.0,0.4,0.0}
\definecolor{darkblue}{rgb}{0.0,0.0,0.4}
\usepackage[bookmarks,linktocpage,colorlinks,
    linkcolor = darkred,
    urlcolor  = darkblue,
    citecolor = darkgreen]{hyperref}
\wocname{EPJ Web of Conferences}
\woctitle{Lattice2017}

\newcommand{\bea}{\begin{eqnarray}}
\newcommand{\ena}{\end{eqnarray}}
\newcommand{\Sb}{\mathcal{S}\:\!}
\newcommand{\DS}{\delta _S}
\newcommand{\dlangle}{\left\langle \kern-.17em \left\langle}
\newcommand{\drangle}{\right\rangle \kern-.17em \right\rangle}
\newcommand{\rhoLLR}{\rho_{\mbox{\tiny LLR}}}
\newcommand{\SMAX}{\Sb_{\mathrm{max}}}
\newcommand{\SMIN}{\Sb_{\mathrm{min}}}
\newcommand{\myend}{\clearpage
\bibliography{lattice2017}
\end{document}
}
\begin{document}
%
\selectlanguage{english}
\rightline{CERN-TH-2017-213}
\title{%
Ergodicity of the LLR method for the Density of States
}
\author{%
  \firstname{Guido} \lastname{Cossu}\inst{1}
  \and
  \firstname{Biagio}  \lastname{Lucini}\inst{2}
  \and
\firstname{Roberto}  \lastname{Pellegrini}\and
\firstname{Antonio} \lastname{Rago}\inst{3,4}\fnsep\thanks{Speaker,
  \email{antonio.rago@plymouth.ac.uk}}
}
\institute{%
School of Physics and Astronomy, The University of Edinburgh,
Edinburgh EH9 3JZ, UK
\and
Department of Mathematics, Swansea University, Swansea SA2 8PP, UK
\and
Centre for Mathematical Sciences, Plymouth University, 
Plymouth, PL4 8AA, United Kingdom
\and
CERN, Theoretical Physics Department,
Geneva, Switzerland
}
\abstract{%
  The LLR method is a novel algorithm that enables us to evaluate the density of states in lattice gauge theory. 
We  present our study of the ergodicity properties of the LLR
algorithm for the model of Yang Mills SU(3). 
We show that the use of the replica exchange method 
alleviates significantly the topological freeze-out that severely affects other algorithms.
}
\maketitle
\section{Introduction}\label{intro}

Monte Carlo (MC) importance sampling simulations are at the basis of
lattice investigations and are one of the most powerful methods to
study non-perturbative phenomena in quantum field theory. Despite
their success, there is a number of known cases where they perform
very poorly, for example in presence of an overlap problem, i.e. when
the measure distribution induces on the update algorithm a poor sample of the portion of phase
space relevant for the evaluation of the observables, or in case of quantities that cannot be expressed as
expectation values over a probability measure like partition functions
and free energies. Over the years many algorithms has been proposed that try to overcome some of these
problems. In this work we investigate the ergodicity properties of a
recently proposed method, the LLR method.

\section{The method}\label{sec-1}

Our method is based on the evaluation of the density of states, and has been recently
proposed in \cite{Langfeld:2012ah,Langfeld:2015fua}.  The density of states $\rho(\Sb)$ is defined via the functional integral
\bea
\rho(\Sb)=\int [D\phi]\,\delta (\mathcal{S}-S[\phi]),
\ena
where $\phi$ is some generic dynamical field variable, $[D\phi]$  is the path
integral over the field configurations and $S[\phi]$ the action of the
Euclidean quantum field theory. From the density of states it is
possible to compute the partition function and expectation values of
energy dependent observables via a simple one dimensional integration:
\bea
Z=\int \rm{d} \Sb\, e^{-\Sb} \rho(\Sb),\\
\langle \mathcal{O} \rangle = \frac{1}{Z} \int \rm{d} \Sb\, e^{-\Sb}
\rho(\Sb)\, \mathcal{O}(\Sb),
\ena
while a bit more of care needs to be payed to evaluate generic
observables (see sect.\ref{method}).

The conceptual difference of this formulation with respect to
classical Montecarlo techniques is that all the path integrals are
evaluated on a micro-canonical ensemble, i.e. at fixed action.

\subsection{The LLR method in a nutshell}\label{method}
The LLR algorithm gives access to a controllable approximation of the
$\rho(\Sb)$ featuring strong convergence properties.
The procedure can be summarised in the steps below\footnote{For a
  more detailed description of the algorithm the reader should refer to \cite{Langfeld:2015fua}.}.

For a given action range, partition the interval between $\SMIN$
  and $\SMAX$ in $N$ subintervals and call  $\Sb_i$ the center action
  value and $\DS=(\SMAX-\SMIN)/N$, for evenly spread actions intervals\footnote{The same procedure 
    works for unevenly distributed intervals with $(\delta _{S})_i=\alpha_i \,\DS$
    and $\sum_{i=1}^{N} \alpha_i/N=1$. }. For each of the intervals evaluate 
\bea
a_i=\left.\frac{d\,\ln
      \rho}{d\Sb}\right|_{\Sb=\Sb_i},
\ena and from the knowledge of the
  coefficients $a_i$ reconstruct the log-linear
  approximation of the density of states as
\bea
\rhoLLR(\Sb)=\rho_0\prod_{i=1}^{k-1}e^{a_i\DS}
{\rm{exp}}(a_k(\Sb-\Sb_k)),\hspace{2cm}\Sb_k\leq \Sb< \Sb_{k+1}\,.
\label{rhollr}
\ena 
It can be shown that our approximation converges to the
correct function: $\displaystyle
\rho(\Sb) = \rhoLLR(\Sb)\, e^{c\,\DS^2} $
“almost everywhere” for $\DS\to0$ (the $\rho(\Sb)$ is supposed to be almost everywhere
$C_2$).

To evaluate the $a_i$ we first define: \hspace{.5cm}
\bea
 \dlangle \Sb -\Sb_i \drangle_{i} (a)
&=& \frac{1}{{\cal N}} \int d\Sb\, e^{\frac{-(\Sb-\Sb_i)^2}{2\DS^2}} \,
\rho(\Sb)\, (\Sb-\Sb_i)\, e^{a \Sb } \\
&=& \frac{1}{{\cal N}} \int [D\phi]\, e^{\frac{-(S[\phi]-\Sb_i)^2}{2\DS^2}} \, (S[\phi]-\Sb_i)\, e^{a S[\phi] } \;,\\
{\cal N} &=& \int d\Sb\, e^{\frac{-(\Sb-\Sb_i)^2}{2\DS^2}} \,
\rho(\Sb)\, e^{a \Sb } =\int [D\phi]\, e^{\frac{-(S[\phi]-\Sb_i)^2}{2\DS^2}} \, e^{a S[\phi] }\;,
\ena
where the double brackets  $\dlangle \dots \drangle$ represent the
insertion of  a strongly localised support function centred around
the $\Sb_i$, in
the present case a gaussian function. 
The $a_i$ are the zeros of the stochastic equation 
\bea
\dlangle \Sb -\Sb_i \drangle_{i} (a_i)=0.
\ena
The stochastic nature of the latter is due to the determination of the
double bracket expectation value, which can be accessed only through Monte
Carlo estimates.  
The numerical solution of stochastic equations can be found using the
iterative Robbins-Monro procedure \cite{robbins1951}.  
This begins with a guessed value $a^{(0)}_i$, which is updated using the
iteration%
\bea
a^{(n+1)}_i = a^{(n)}_i -c_n\dlangle \Sb-\Sb_i\drangle_i(a^{(n)}_i)\hspace{1cm} {\rm with}\;
\sum_{n=0}^\infty c_n=\infty \; {\rm and} \;\sum_{n=0}^\infty c_n^2<\infty,
\ena
and converges in $L_2$, and hence in probability, to root of the equation:
\bea
\lim_{n\to\infty}a^{(n)}_i=a_i,
\ena
for more details see \cite{robbins1951}. 

Once the $a$’s are computed the density of states is reconstructed using
eq.\ref{rhollr} with error-bars determined using bootstrap resampling.

A few remarks are in order: the $\rhoLLR$ is an approximation of the
correct density of state $O(\DS^2)$  and it shows exponential error
suppression: the relative approximation error does not depend on the
magnitude, as a consequence this method works over several orders of magnitude.
Once the $a_i$ have been evaluated, energy observables can be evaluated through direct integration of the
density of states while generic observables can be evaluated through 
\bea
\langle \mathcal{B} \rangle  = \frac{1}{Z} \, \sum _i 
\DS \;\rhoLLR \left(\Sb_i\right) e^{-a_i \Sb_i}\;
\dlangle \mathcal{B}[\phi] \, e^{ -(a_i+1) S[\phi ]} \drangle + O(\DS^2) ,
\ena
and also in this case the convergence in $\DS$ is $O(\DS^2)$.

Note also that being the support function analytic this method is
amenable to HMC simulation at practically no additional coding and
computational cost, see \cite{Pellegrini:2017iuy}.

To summarise for each step of our iteration we have to evaluate our
observables over configurations distributed according to the weight
\bea
W(\phi,\Sb_i,a)\propto e^{a S[\phi]-\frac{(S[\phi]-\Sb_i)^2}{2\DS^2}}=e^{U[a,\phi,\Sb_i]}.
\ena
Once the Robins-Monro procedure has converged we can reconstruct the observables of the full theory by an
appropriate unidimensional integration. It is however clear that the
strong localisation in action values imposed by the presence of the
gaussian could slow down the update dynamic. Indeed the probability of visiting states with action far from the
peak of the gaussian will be parametrically small in $\DS$ and this will lead to a slow
dynamic of the Markov Chain.

The proposed solution is to simultaneously simulate multiple
overlapping intervals with fixed central action and periodically
propose a swap of the configurations belonging to pairs of them with
probability:

\bea
P_{\rm swap}=\min(1,\exp(U[a_1,\phi^{(1)},\Sb_1] +
U[a_2,\phi^{(2)},\Sb_2]
- U[a_2,\phi^{(1)},\Sb_2]
- U[a_1,\phi^{(2)},\Sb_1])).
\ena

Such step preserves the detailed balance of action of the entire system
and hence the resulting algorithm is still ergodic.
Subsequent exchanges allow any configuration sequence to
travel through all the action intervals, hence overcoming any
potential action barrier.
In the following we will refer to this method as replica\footnote{With
    replica we want to signal the part of the simulation related to the
    constrained evolution around a specific $\Sb_i$.}  exchange, however
  it has to be noted that in literature it has also been referred to
  as umbrella sampling or parallel tempering.
This method has been applied to the study of systems with
strong metastabilities like the $q$-state Potts model at large $q$
\cite{Lucini:2016fid}. In this work we want to investigate the behaviour of the
algorithm for theories with  topological sectors by studying 
autocorrelation time of observables sensitive to these sectors.
\section{The model}
\label{themodel}
We studied a pure SU(3) Yang Mills model in $d=4$ with Wilson
action.
To simulate the model we used an HMC update, with Molecular Dynamics
trajectory length $\tau=1$, we used a $2^{\rm nd}$ order Omelyan
integrator tuned to obtain an acceptance $\sim 98\%$ for all our runs. 
For the choice of the MC parameter and a  direct comparison of our data we refer
to \cite{Schaefer:2010hu}. Our simulations were performed for a range of
action values corresponding to $5.789\leq\beta\leq6.2$ or
  $0.140fm\geq a \geq 0.068fm$ for fixed number of lattice points
  $V=16^4$ hence with a extension ranging from $2.2fm$ to $1.1fm$.

The observables that we measured are:
\begin{itemize}
\item Action observables: observables that can be written as a function of
the only action (these are a byproduct of the LLR method).
\item The flowed action density $E$ defined through the average of the plaquettes
evaluated over the link evolved with the Wilson Flow at flow time $t$
and its more symmetric clover definition $E_{\rm clover}$ \cite{Luscher:2010iy}.
\item The topological charge, using a bosonic estimator computed along the Wilson flow:
\bea
  Q_t = -\frac{a^4}{32\pi^2} \sum_x \epsilon_{\mu\nu\rho\sigma}
  \mathrm{tr} \left[G_{t,\mu\nu}(x) G_{t,\rho\sigma}(x)\right]\, ,
\ena
where $G_{t,\mu\nu}(x)$ is again the clover term for the field strength tensor
on the lattice constructed from the gauge links at flow time
$t$~\cite{Luscher:2011kk}.
\end{itemize}

\begin{figure}[thb] 
\centerline{\includegraphics[width=.5\textwidth]{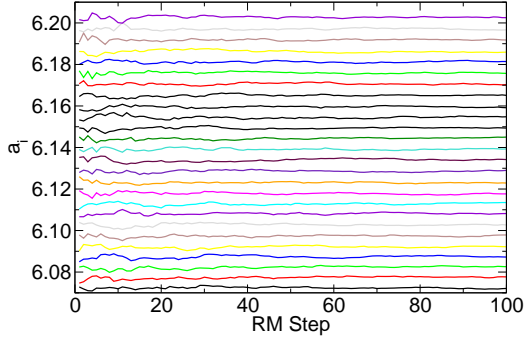}}
\caption{Determination of the $a_i$ coefficient along the Robbins Monro
  evolution for all the action intervals }
  \label{fig-ai}
\end{figure}

For all the above observables we use the Madras-Sokal definition
of the autocorrelation time~\cite{Madras:1988ei}, implemented
according to Refs.~\cite{Wolff:2003sm,Luscher:2005rx}.
$\bar{\Gamma}(\xi)$ is the
autocorrelation as function of the time lag $\xi$ between measurements:
\bea
  \bar{\Gamma}(\xi) = \frac{1}{N-\xi} \sum_{i=1}^{N-\xi} \left(\mathcal{B}_i - \bar{\mathcal{B}}\right)
  \left(\mathcal{B}_{i+\xi} - \bar{\mathcal{B}}\right)\, ,
\ena
where $N$ is total number of measurements, $\mathcal{B}_i$ denotes the i-th
measurement of the observable $\mathcal{B}$, and $\bar{\mathcal{B}}$ its average.
The integrated autocorrelation time
$\tau_\mathrm{int}$ is defined as:
\bea
  \tau_\mathrm{int} =  \frac12 + \sum_{\xi=1}^{W}
  \bar{\Gamma}(\xi)/\bar{\Gamma}(0)\, ,
\ena
where $W$ is the size of the Madras-Sokal window. 
\subsection{Tuning of the replica exchange method}
\begin{figure}[thb] 
  \centering
\includegraphics[width=.70\textwidth]{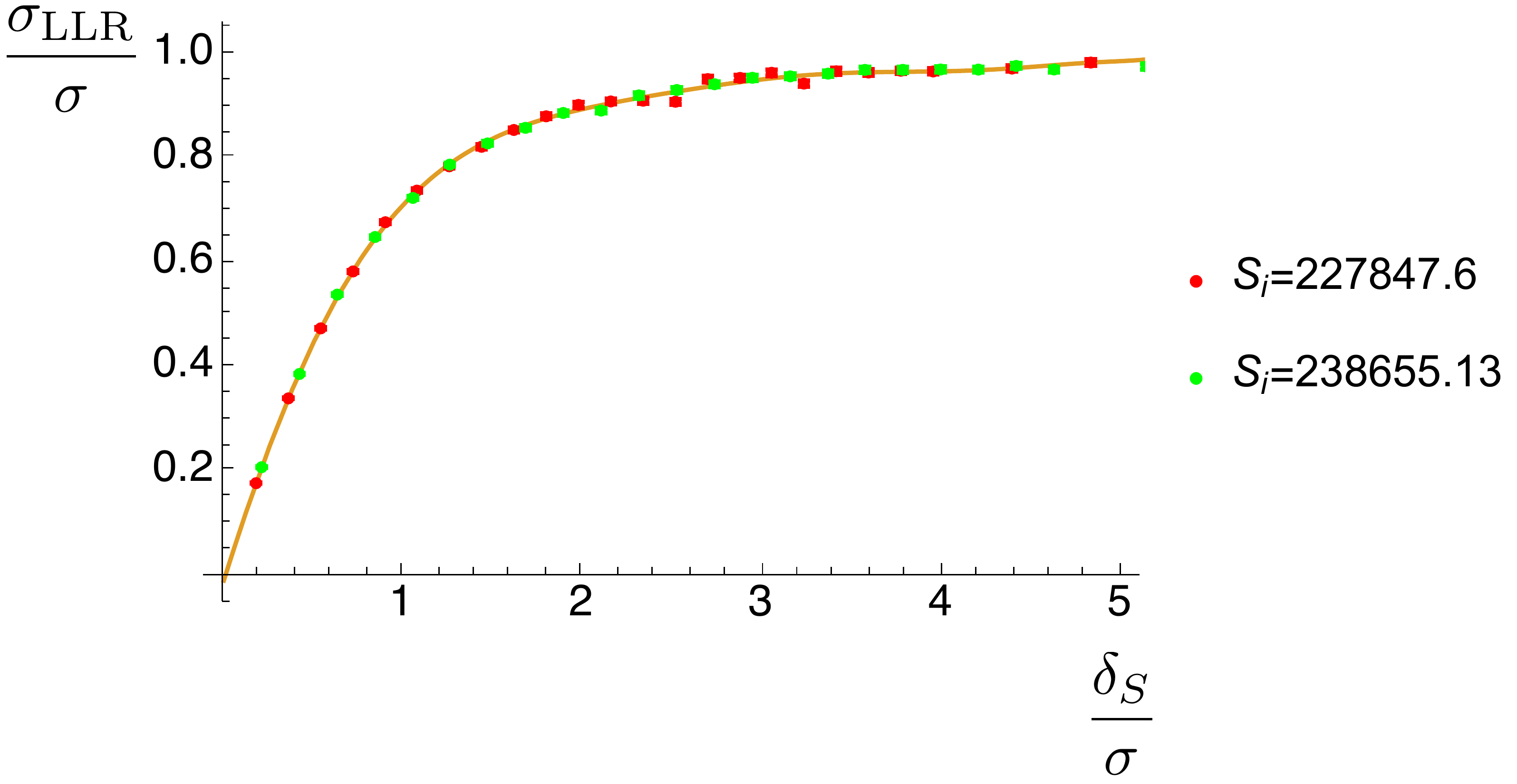}
\caption{Behaviour of the variance of the constrained action versus
  the width of the support function normalized to the variance of the
  unconstrained action }
  \label{fig-1}
\end{figure}
\begin{figure}[thb] 
  \centering
\includegraphics[width=.7\textwidth]{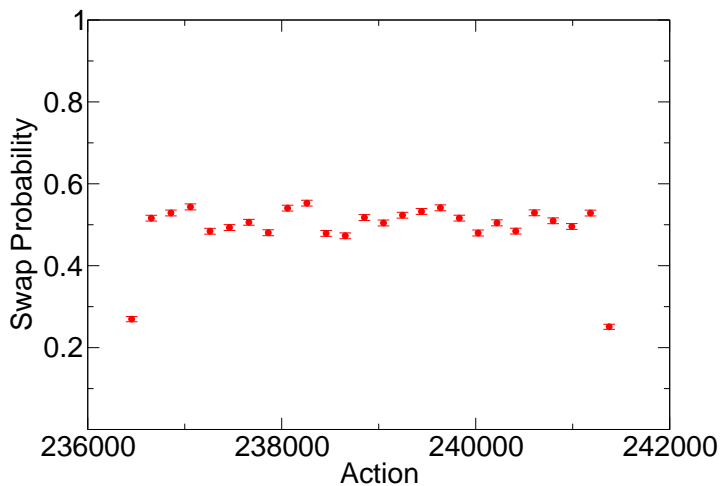}
\caption{Swap probability as function of the central action }
  \label{fig-2}
\end{figure}

The swapping probability of two replicas will depend on the size of
their overlapping probability distributions, hence typically it will depend on
$\Sb_i$ and on the fluctuation size of the constrained evolution
$\sigma_{\rm LLR}=\sqrt{\dlangle \Sb^2\drangle_i - \dlangle
  \Sb\drangle^2_i}$. In order to tune the swapping probability among
different replicas we noted that it is possible to relate the $\sigma_{\rm
  LLR}$ to the width of the support function ($\DS$) and to the
variance of the unconstrained evolution $\sigma$ for identical action values.
This study is reported in figure \ref{fig-1}  where the asymptotic behaviour both for small $\DS$ ($\sigma_{\rm}\simeq\DS$) and large $\DS$ ($\sigma_{\rm
  LLR}\simeq\sigma$) can be clearly highlighted. An additional feature
of this study is the independence within our numerical precision of
the scaling function with respect to the central action value.
Thanks to this feature, it becomes possible, with the only knowledge of
the variance of the unconstrained action  as function of $\beta$, to
generate a sequence of $\Sb_i$ and $(\DS)_i$ for which the overlapping
spread of action among neighbouring replicas is equal. If we define the
swap probability as the probability for fixed energy value to swap configurations 
with any other replica, it is possible to show that the tuning just
described gives rise to a flat swap probability among all the
replicas. By changing the size of the overlapping region it is
possible to tune the swapping probability to the desired value, in figure
\ref{fig-2} the case of probability $\sim50\%$.

A final note is in order: the Markov Chain process defined in this way
is still distributed according the detailed balance of energy 
when considering the whole set of replicas, as such the
autocorrelation time can be measured only on the 
observables reweighed over all the replicas together and not on a
single replica at a time.
\section{Results}
We have performed an extensive investigation of the properties of our
algorithm and compared our findings against the results of traditional HMC
simulations for all the observables described in
sec.\ref{themodel}, together with their autocorrelation
times. 
For a rough cost budget we have generated $0.7 \cdot 10^5$
configurations per replica and divided the action interval over 26
replicas, while for each traditional simulation we have generated $1.1 \cdot
10^5$ configurations. We produced estimates of our observables only in
an internal region defined such that the reweighed observables could
have a contribution from the boundary $10^{-4}$ times smaller than the
central contribution.
All our estimates agree within the statistical error with their
reference comparisons, and depending on the observables the statistical
error associated with the observables measured with the LLR approach
is at least a factor 2 smaller then traditional
simulations. More importantly the
autocorrelation time of all the investigated observables, once the replica exchange is taken into account,
is smaller than the equivalent traditional simulation, signalling
a good ergodicity property of the algorithm.

\begin{figure}[thb] 
\centerline{\includegraphics[width=.5\textwidth]{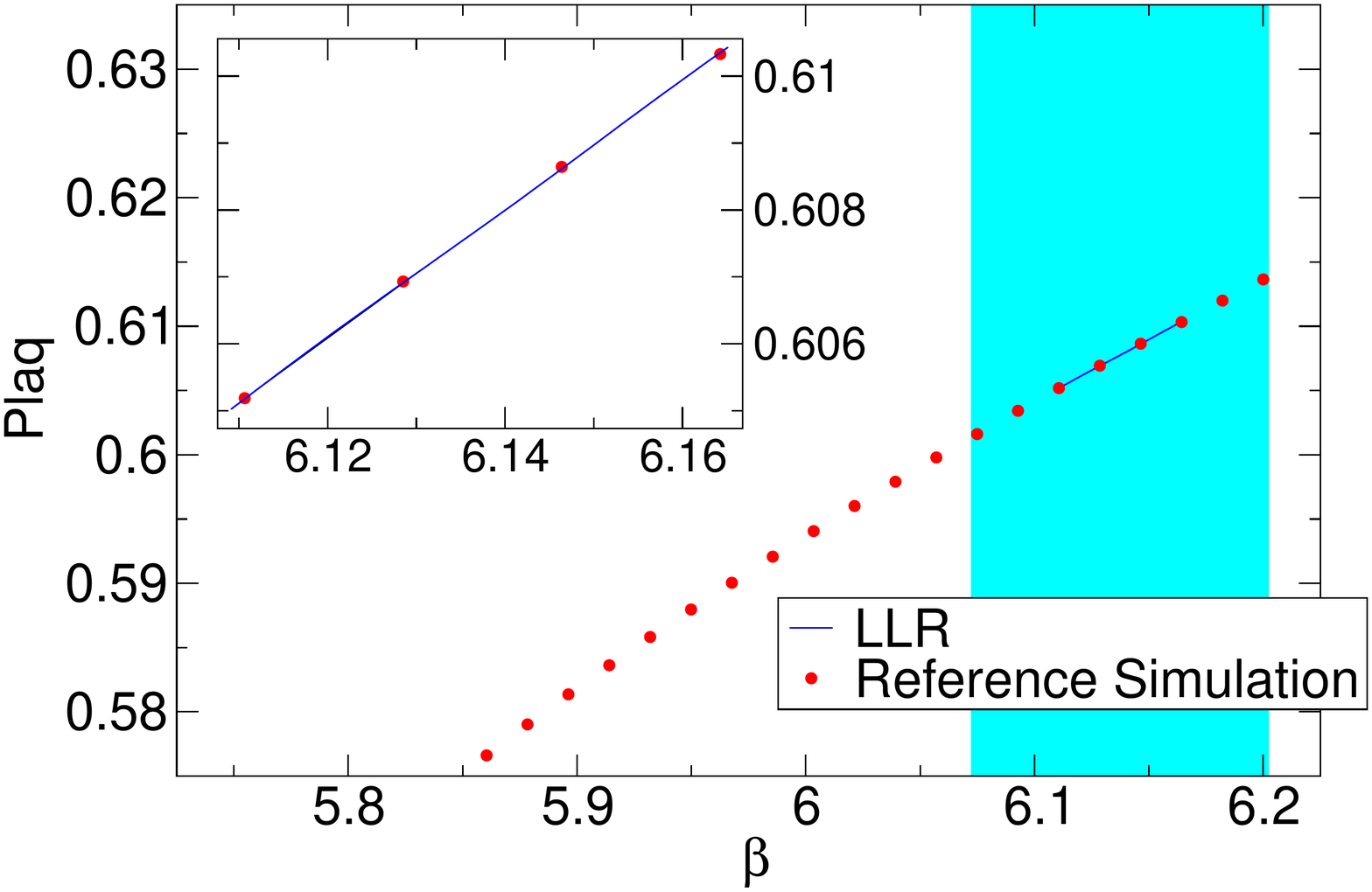}\hspace{.2cm}\includegraphics[width=.5\textwidth]{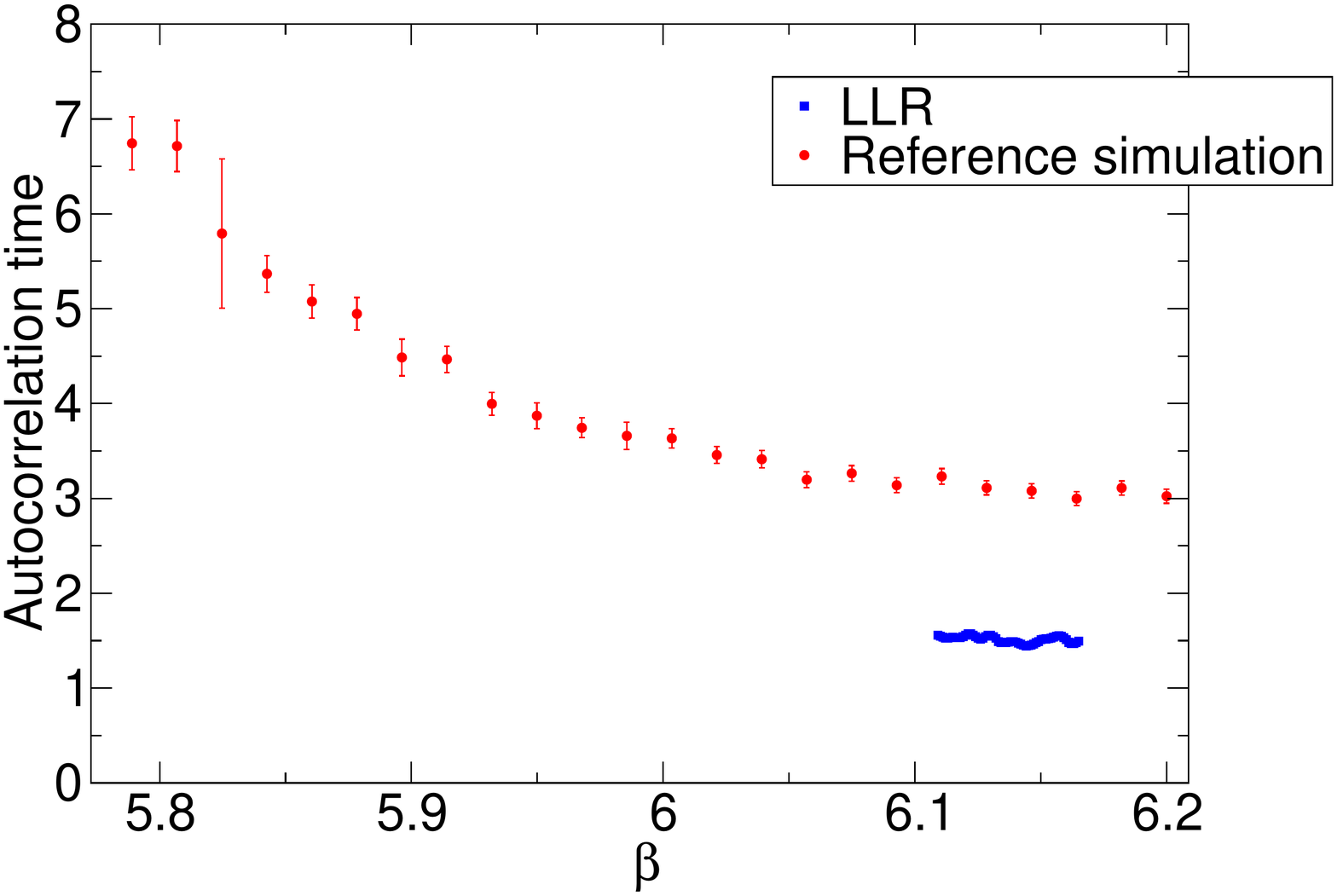}}
\caption{Plaquette and its autocorrelation time as function of $\beta$ at
fixed lattice volume}
  \label{fig-3}
\end{figure}

\begin{figure}[thb] 
\centerline{\includegraphics[width=.5\textwidth]{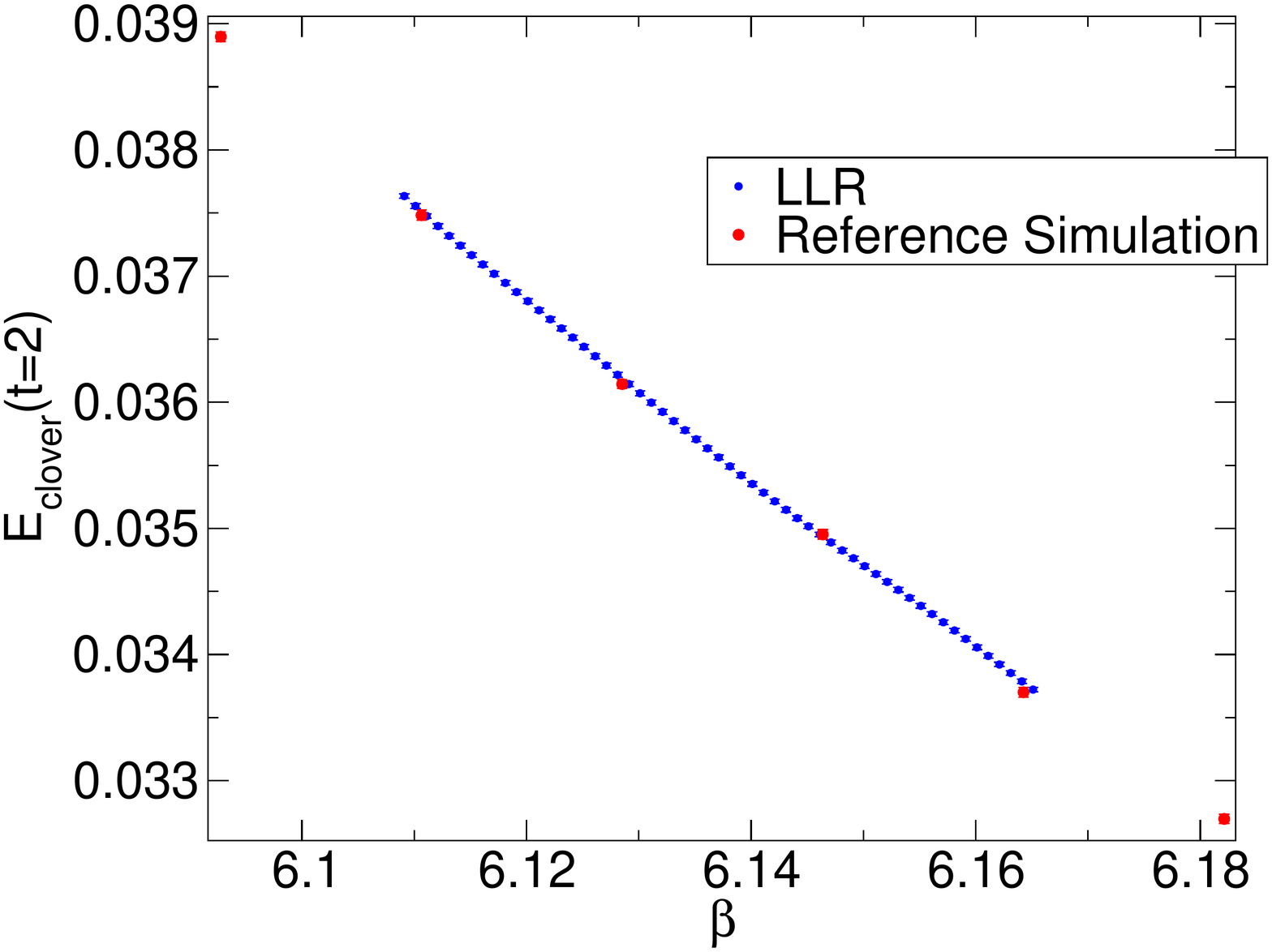}\hspace{.2cm}\includegraphics[width=.5\textwidth]{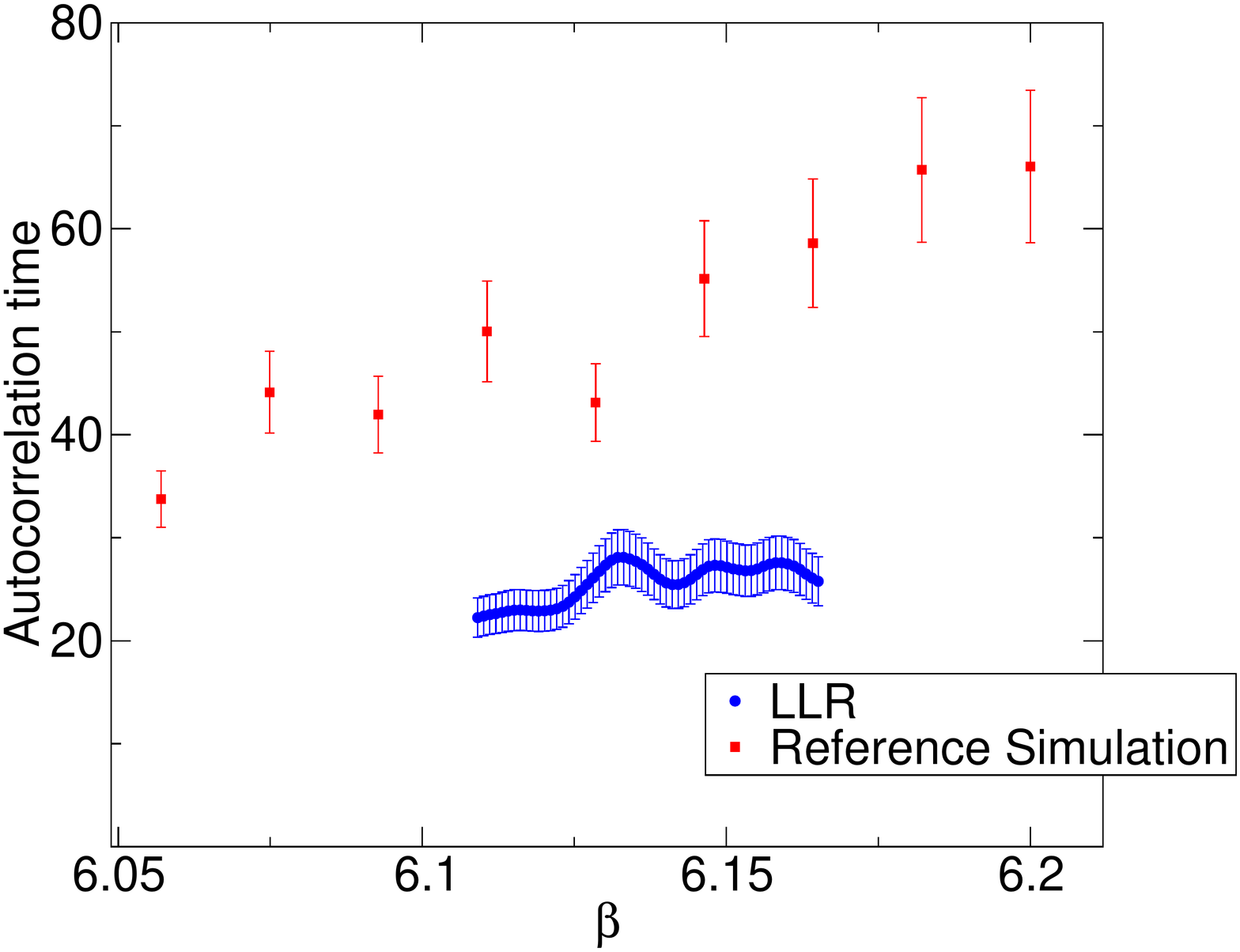}}
\caption{Clover energy at flow time $t=2$ and its autocorrelation time as
  function of $\beta$ at fixed lattice volume }
  \label{fig-4}
\end{figure}

\begin{figure}[thb] 
\centerline{\includegraphics[width=.5\textwidth]{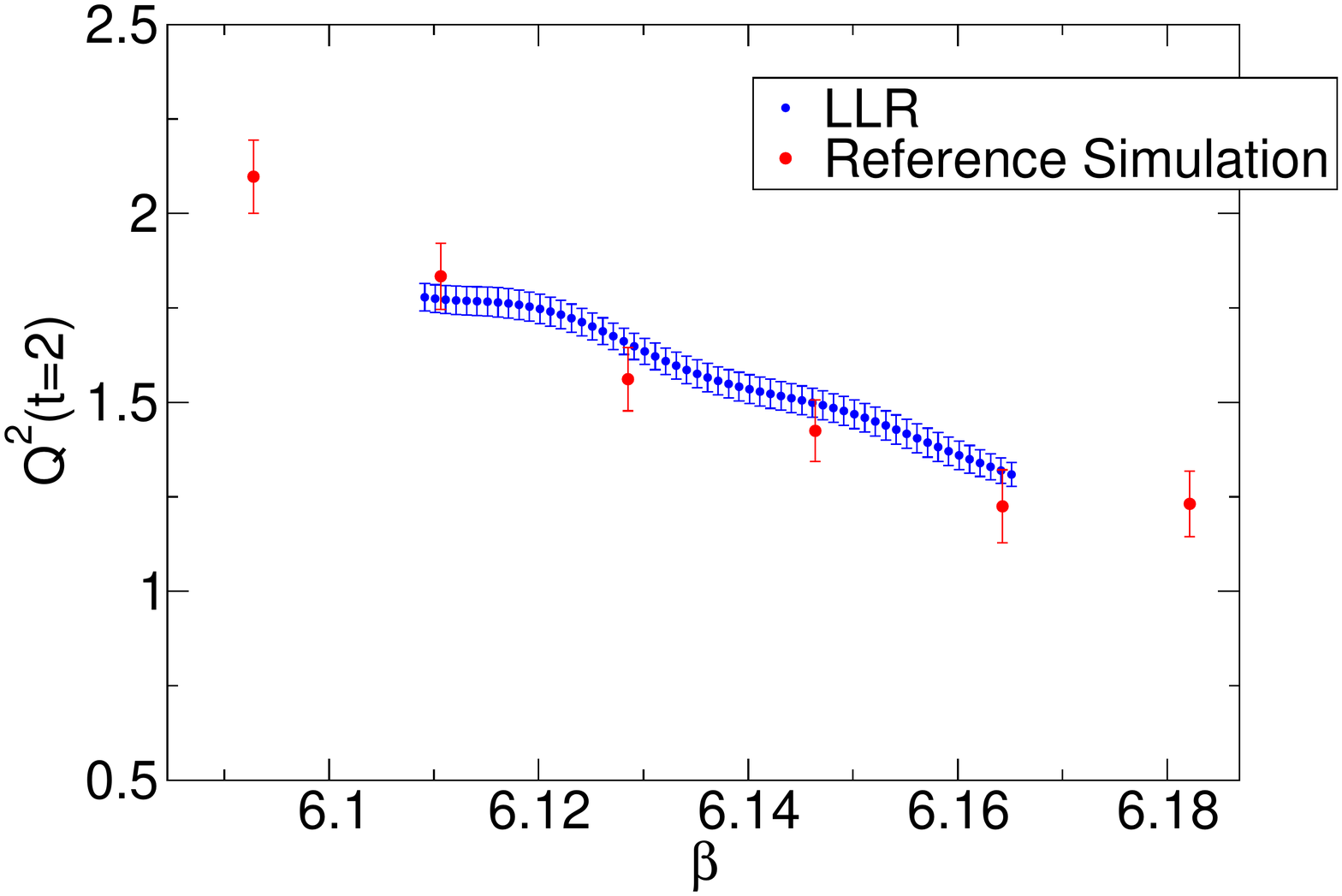}\hspace{.2cm}\includegraphics[width=.5\textwidth]{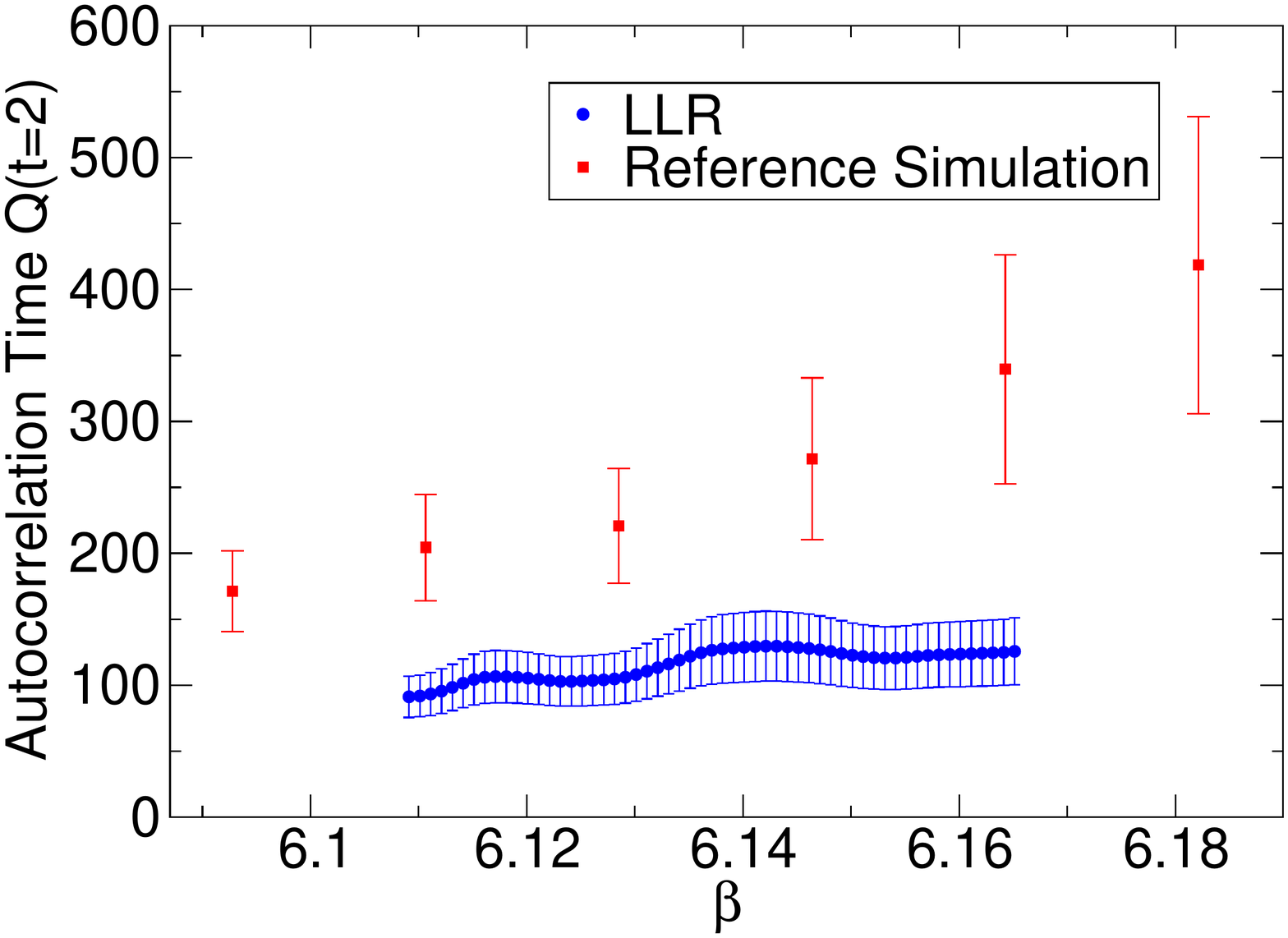}}
\caption{Topological susceptibility at flow time $t=2$ and the
  autocorrelation time of the topological charge as
  function of $\beta$ at fixed lattice volume}
  \label{fig-5}
\end{figure}
\section{Conclusions}
In this work we have investigated the ergodicity properties of the LLR
method to evaluate the density of state. We focused in our study on
the SU(3) pure Yang Mills theory, to test the behaviour of our method
with theories that possess topological vacua. We measured observables sensitive
to the topological structure  alongside other less sensitive
observables, and compared our findings with
equivalent investigation done with traditional techniques.
We have shown that the use of a replica exchange technique alleviates the
problem of slow update dynamic evolution also in presence of the topological vacua.
We have shown that the LLR method is capable of comparable precision
in all the estimates at a comparable resource cost once the
autocorrelation is taken into account. 
Simulation are still ongoing to investigate the scaling properties of
this algorithm towards the continuum limit.

\section{Acknowledgments}
GC is supported by the STFC Consolidated Grant ST/L000458/1. BL is supported
  in part by the Royal Society and the Wolfson Foundation and the STFC
  Consolidated Grants ST/L000369/1 ST/P00055X/1.  AR is supported by the
  STFC Consolidated Grants ST/L000350/1 and  ST/P000479/1.
The numerical computations have been carried out using resources from the HPCC Plymouth and the Supercomputing Wales (supported by the ERDF through the WEFO, which is part of the Welsh Government).
\bibliography{lattice2017}

\end{document}